\documentclass[12pt]{iopart}
\usepackage{graphicx}

\begin{document}

\title{Competition and evolution in restricted space}

\author{F. L. Forgerini$^{1}$, N. Crokidakis$^{2}$}

\address{
$^{1}$ ISB - Universidade Federal do Amazonas \\
69460-000 \hspace{5mm} Coari - AM \hspace{5mm} Brazil\\
$^{2}$ Instituto de F\'isica, Universidade Federal Fluminense \\
24210-340 \hspace{5mm} Niter\'oi - RJ \hspace{5mm} Brazil}

\ead{fabricio$\_$forgerini@ufam.edu.br, nuno@if.uff.br}

\begin{abstract}
\noindent
We study the competition and the evolution of nodes embedded in Euclidean restricted spaces. The population evolves by a branching process in which new nodes are generated when up to two new nodes are attached to the previous ones at each time unit. The competition in the population is introduced by considering the effect of overcrowding of nodes in the embedding space. The branching process is suppressed if the newborn node is closer than a distance $\xi$ of the previous nodes. This rule may be relevant to describe a competition for resources, limiting the density of individuals and therefore the total population. This results in an exponential growth in the initial period, and, after some crossover time, approaching some limiting value. Our results show that the competition among the nodes associated with geometric restrictions can even, for certain conditions, lead the entire population to extinction.
\end{abstract}

%Keywords: Population biology, Complex systems, Branching process, Computer modeling and simulation.
\maketitle

%%%%%%%%%%%%%%%%%%%%%%%%%%%%%%%%%%%%%%%%%%%%%%%%%%%%%%%%%%%%%%
\section{Introduction}

A growing tree-like network can model different processes such as a technological or biological systems represented by a set of nodes, where each element in the network can create new elements. Innovation and discovery \cite{paczuski}, artistic expression and culture \cite{Cowlishaw}, language structures \cite{paulo_murilo,sergey1} and the evolution of life \cite{koonin,moret} can naturally be represented by a branching process in a tree \cite{huson} describing a wide range of real-life processes and phenomena \cite{livro:DM,livro:enc_comp_sys_branching,gray,Dorogovtsev:dkm08,Dorogovtsev:dgm2008,Albert:ab02,Newman:n03}.

The general branching process is defined mathematicaly as a set of objects (nodes) that do not interact and, at each time step, each object can give rise to new objects. In contrast, interacting branching processes are much more interesting and difficult for analysis \cite{paczuski}. A generalized tree with one (or more) ancestor(s) have been used to depict evolutionary relationships between interacting nodes such as genes, species, cultures. Besides the interaction among nodes, one can consider spatially embedded nodes. The evolution of networks embedded in metric spaces have been attracted much attention \cite{Dall:dc02,Kleinberg:k00,Carmi:ccs09,Cartozo:cd09,Boguna:bk09,Krioukov:kpk10}.

In this work we study the evolution of a population, i.e., the number of nodes in the network, influenced by the interaction among existing nodes and confined to a limited area, representing a competition of individuals for resources. We assume that the growing tree is embedded in a metric space and we consider that spatially close nodes, previously placed in the network, will suppress their ability to born new nodes. In other words, overcrowding of nodes will drain the resources and supress the offspring. In our model each node lives for three generations.

The evolution of the population of nodes is actually determined by two parameters: the minimum distance between any pair of nodes $\xi$, and the area in which the network is embedded, namely the linear size of the area, $L$. For simplicity, we assume that this area does not change in time. The population evolves in two different regimes. At the initial generations (time steps), one can see an exponential evolution, followed by a saturation regime, after a crossover time.

In the saturation regime, the size of the network will finally approach some limiting value. The network has even a chance to extinguish if at some moment all its nodes occur in a small area. We investigated this possibility of complete extinction. The term extinction for our model implies the end of evolution and the absence of new generations. The interaction among the nodes inside the radius is defined by a parameter $\xi$ and the value of $L$ regulates the population dynamics. Our results show that, under certain conditions, the entire population can be led to extinction.

This paper is organized as follows. In Sec.~2 we present our model details and obtain simple estimates for its growth. In Sec.~3 we describe the populational evolution. The possibility of extinction for the model embedded in a bounded space is discussed in Sec.~4, and, finally, in Sec.~V, we summarize the results and present our conclusions.

%#######################################################

\section{The Model}
\label{model}

\begin{figure}%[!ht]
\includegraphics[width=\linewidth]{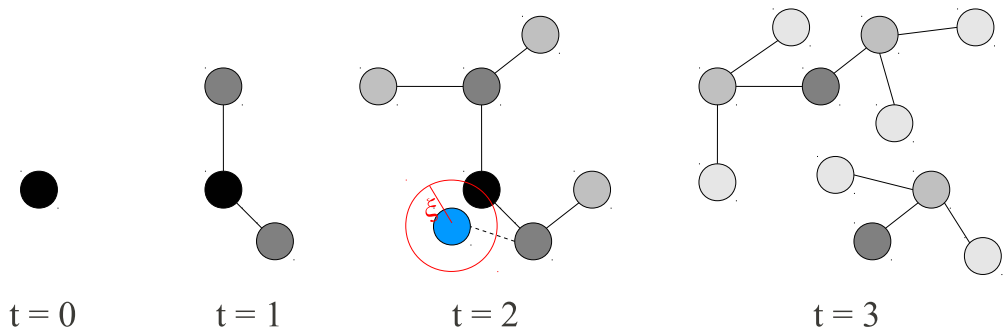}
\caption{Schematically representation of the growing process in the network. Starting from the root node (black circle) two new nodes are added in $t=1$ and new attempts are made each time step. In $t=2$, one can see a refused attempt (blue circle) due to the proximity to other nodes (closer than a distance $\xi$). In $t=3$, the oldest node is removed and new nodes are created.}
\label{fig1}
\end{figure}

In our model, the population consists of interacting nodes spatially separated by some distance. We start our process from a single root node at time $t=0$, as one can see in Fig.~\ref{fig1}. The single root node (black circle in Fig.~\ref{fig1}), can branch to produce up two new daughter nodes (dark gray circles) at future generation, i.e., at the next time step. The position of each new node is randomly chosen inside a circle with a given \textit{radius} ($0 \le radius \le 1$) centered in the parents' positions. The attempt to add a newborn node is refused in the case the chosen position is closer than at distance $\xi$ from other nodes.

The attempt to generate offsprings takes place at the next time step after the introduction of a new node in the network and each node can produce daughter nodes only at this time. At the next time step, after three generations, the node is removed from the network.

%%%%%%%%%%%%%%%%%%%%%%%%%%%%%%%%%%%%%%%%%%%%%%%%%%%%%%%%%%%%%
%%%%%%%%%%%%%%%%%%%%%%%%%%%%%%%%%%%%%%%%%%%%%%%%%%%%%%%%%%%%%

At each time step, each of the nodes previously introduced, attempts to branch, so at each time step a new generation of nodes is born. The nodes are chosen uniformly at random one by one and during a unit of time we update the entire network. The total area of the system is limited, considering that it is natural the introduction of a spatial restriction into the model. The first node is settled as the origin of the space and from the origin we set a maximum length for each spatial coordinate of a two-dimensional space. In other words, the geometric position of each node in the network, for our model, is restricted in the range $-L/2 \le x \le L/2$, $-L/2 \le y \le L/2$. The linear size of the area, $L$, is introduced as a parameter of the model and we assume that this area does not change in time. In our simulations we used open boundary conditions.

%#######################################################

\section{Population size evolution}
\label{size}

\begin{figure}%[!ht]
\includegraphics[width=0.6\linewidth]{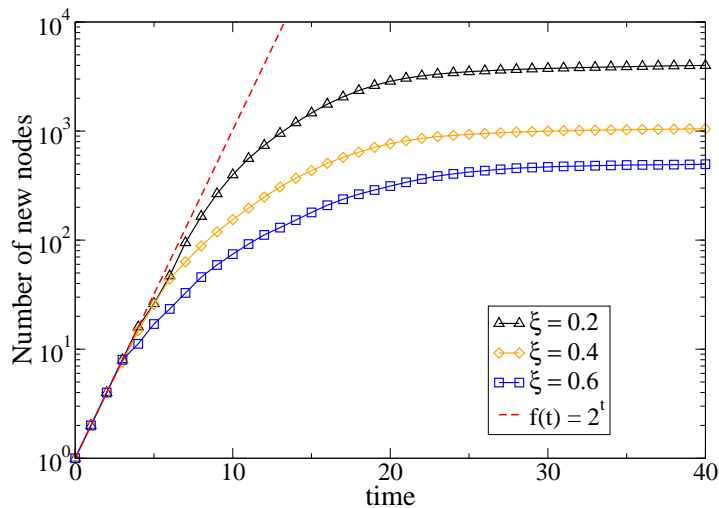}
\caption{Time evolution of the number of new nodes for different values of $\xi$. The behavior for the initial time steps, $f(t)\sim 2^{t}$, is also exhibited. Data are averaged over 50 samples.}
\label{fig2}
\end{figure}

If one lets the population dynamics evolve embedded in a infinitely large system ($L~\rightarrow~\infty$), the population always increase in size. The number of new nodes grows very fast as $N = 2^t$ for initial times, and, after certain crossover time $t_{\times}$, the growth is slower than exponential, as one can see in the Fig.~\ref{fig2}.

%%%%%%%%%%%%%%%%%%%%%%%%%%%%%%%%%%%%%%%%%%%%%%%%%%%%%%%%%%%%%
%%%%%%%%%%%%%%%%%%%%%%%%%%%%%%%%%%%%%%%%%%%%%%%%%%%%%%%%%%%%%

At this regime the total population as function of the time is $P(t) \sim (\frac{t}{\xi})^2$, for $t$ greater than $t_{\times}$. We can estimate, very roughly, $t_{\times}$ from $N(t < t_{\times})=2^{t}$ and $N(t > t_{\times})\sim (\frac{t}{\xi})^2$, we have
\begin{equation}
2^{t_{\times}} \sim (t_{\times}/\xi)^2 ,
\label{e1}
\end{equation}
which leads to the estimate
\begin{equation}
t_{\times} \sim \frac{2}{\ln2}\ln\Bigl(\frac{1}{\xi}\Bigr) ,
\label{e2}
\end{equation}
at small $\xi$. Our numerical results are considering that $t \gg t_{\times}$, for the estimates of the total population in the saturation regime. We should emphasize that in our model the population is confined into a limited area and it is not possible to grow indefinitely.

\begin{figure}[t]
 \vspace{0.2cm}
 \includegraphics[width=0.6\linewidth]{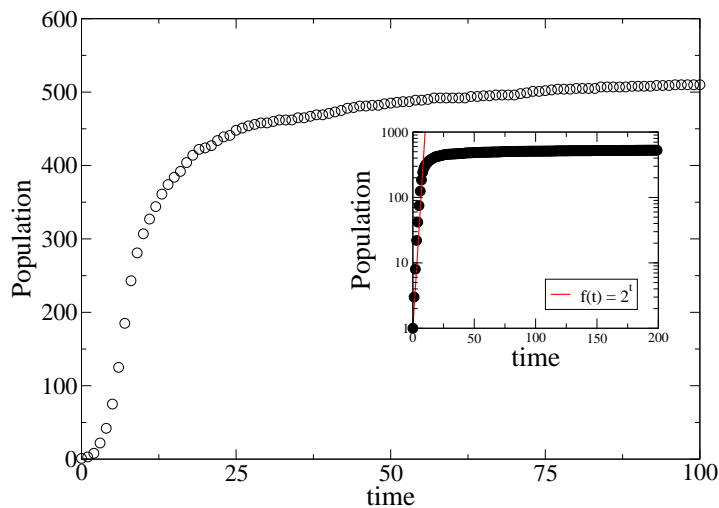}
 \caption{Time evolution of a single realization of the population dynamics, for $L = 10$ and $\xi = 0.2$. The inset shows that even for one sample, the population reaches a constant value after some time.}
 \label{fig3}
\end{figure}

The general result of our simulations for this model is exhibited in Fig.~\ref{fig3}, where we consider a two-dimensional space and a sufficiently small value of $\xi$, in comparison with the linear size of the system $L$. Initially, the population grows exponentially and, after certain crossover time, one can see that the population reach an steady state. After the crossover time, $P(t)$ is nearly constant. The maximum value of the population is $P_{max}=(\frac{L}{\xi})^2$, since we are considering a two-dimensional space for the simulations. The growth of the populational density of \textit{Paramecium} in laboratory, for instance, is reported to have the same behavior of our model~\cite{Vandermeer}.

One can see that in the saturation regime, the total population is smaller than $P_{max}$. This is due to the fact that the interaction among the nodes does not allow that each possible offspring may be created at some generation, keeping the total population below this limit.

%%%%%%%%%%%%%%%%%%%%%%%%%%%%%%%%%%%%%%%%%%%%%%%%%%%%%%%%%%%%%
%%%%%%%%%%%%%%%%%%%%%%%%%%%%%%%%%%%%%%%%%%%%%%%%%%%%%%%%%%%%%

%#######################################################

\section{Extinction}
\label{extinction}

For a small population, the possibility of extinction is higher. The network has even a chance to extinguish if the offspring created are too few and, at some moment, all its nodes occur in a small area. It is a well known characteristic that the smaller a population, the more susceptible it is to extinction by various causes \cite{Sznajd-Weron}.

\begin{figure}[htb]
 \vspace{0.2cm}
 \includegraphics[width=0.6\linewidth]{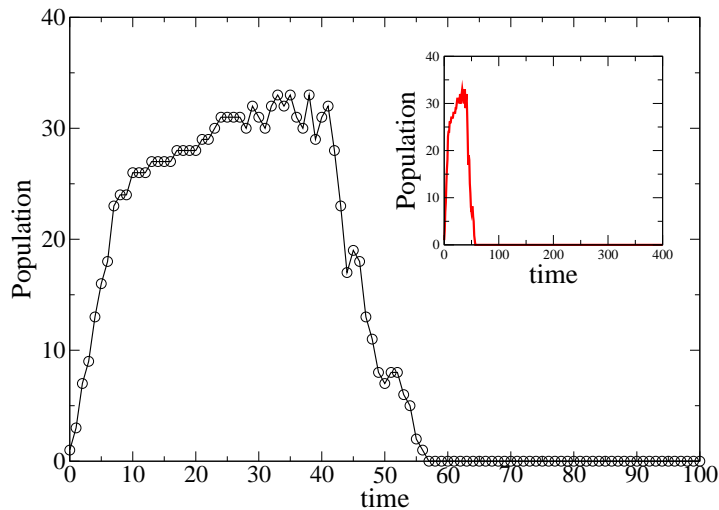}
 \caption{Evolution of the population for a single realization of our model. The population grows for the initial time steps, but it can also decrease and, after some time, extinguish. This behavior is maintained for long times, as one can see in the inset. The parameters are $\xi=0.9$ and $L=1$.}
 \label{fig4}
\end{figure}

Figure~\ref{fig4} demonstrates an example of the evolution of the population, which in this case has $\xi=0.9$ and $L=1$. The population rapidly increase and the system enters in the fluctuation regime, for which the population fluctuates around a mean value for a few generations. After some time, the population decreases and is extinguished.

%%%%%%%%%%%%%%%%%%%%%%%%%%%%%%%%%%%%%%%%%%%%%%%%%%%%%%%%%%%%%
%%%%%%%%%%%%%%%%%%%%%%%%%%%%%%%%%%%%%%%%%%%%%%%%%%%%%%%%%%%%%

The picture which we observe agrees with traditional views on extinction processes which show ``relatively long periods of stability alternating with short-lived extinction events'' (D.~M.~Raup) \cite{raup}. The competition for resources combined with the restricted space limits the total population and, for some values of the parameters in the model, the population should finally extinct. In real situations, extinction may require external factors, an environmental stress~\cite{newman} or an internal mechanism, such as mutation \cite{sibani}. One can see an example in the case of extinction of reindeer population in St. Matthew Island~\cite{klein}. The U.S. Coast Guard released 29 reindeer on the island during World War II, in 1944. The reindeer population grows exponentially and, in 1963, it was about 6,000 animals on the island, 47 reindeer per square mile. The overpopulation, limited food supply and the exceptionally severe winter of 1963-1964 significantly affect future offspring. The reindeer population of St. Matthew Island drops to 42 animals in 1966 and dies off by the 1980s.

This kind of extinction may also occur in branching annihilating random walks and other related processes studied in Refs. \cite{takayasu,jensen}, in which the random processes play the role of an external factor, internal mechanism or an environmental stress that may lead to extinction.

In our model, if we choose a large enough value of the parameter $\xi$, the population will be small and the number of new nodes after some generations can decrease and, sometimes, vanishes. When the nodes' competition increases, the population may decays or even vanishes, as we can see by the rapidly decreasing of the new nodes in Fig.~\ref{fig4}.

\begin{figure}[t]
 \vspace{0.2cm}
 \includegraphics[width=0.6\linewidth]{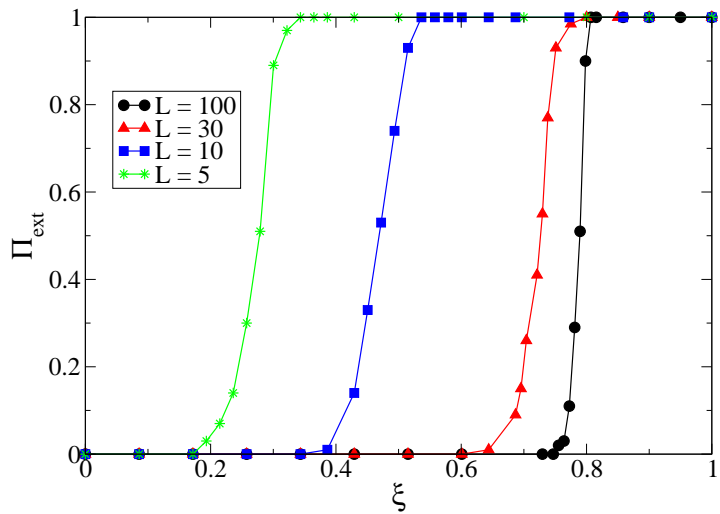}
 \caption{Probability of extinction within $10^{6}$ generations versus $\xi$ for 100 samples and different values of linear size L of the system. }
 \label{fig5}
\end{figure}

We investigated the state of the branching process after a long period of time, $t_{observation}=10^5$ generations (i.e., time steps). For the case when $\xi \rightarrow 1$ the population always dies off, since no offspring is allowed, for any value of $L$. We simulated our model for $100$ different samples and for various values of $L$ and $\xi$. From this data we obtain the probability of extinction $\Pi_{ext}$, i.e., the fraction of samples in which the population dies off before the 10$^{6th}$ generation, for different L, as one can see in Fig.~\ref{fig5}.

%%%%%%%%%%%%%%%%%%%%%%%%%%%%%%%%%%%%%%%%%%%%%%%%%%%%%%%%%%%%%%%%%%%

%%%%%%%%%%%%%%%%%%%%%%%%%%%%%%%%%%%%%%%%%%%%%%%%%%%%%%%%%%%%%%%%%%%

\begin{figure}[htb]
 \vspace{0.2cm}
 \includegraphics[width=0.6\linewidth]{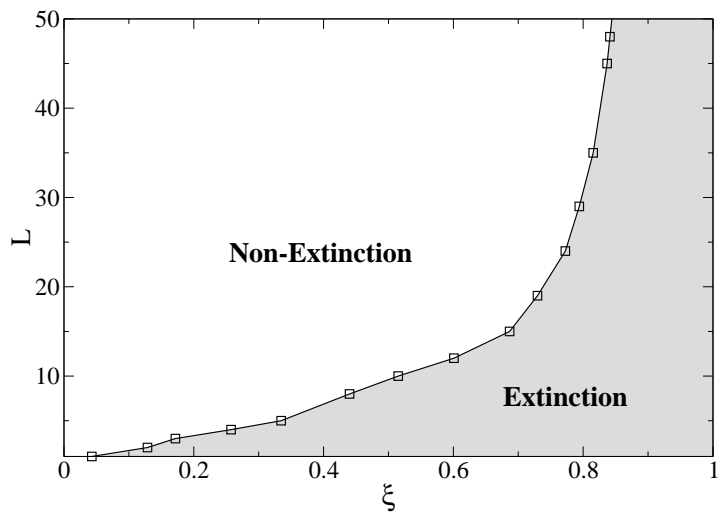}
 \caption{Diagram on the L versus $\xi$ plane, where one can see extinction and non-extinction regions.}
 \label{fig6}
\end{figure}

In Fig.~\ref{fig6} one can see a diagram where an extinction (below the curve) and non-extinction (above the curve) regions are shown. Each point is Fig.~\ref{fig6} is defined as follows. For a given value of $L$ and considering $10^{6}$ generations, the value of $\xi$ for which the probability of extinction goes to one defines one point in the graphic of Fig.~\ref{fig6}. Our results show that high values of populational density, represented in our model for small $L$ and large $\xi$, can be lead to extinction. This picture is a different representation of the probability of extinction in which we are considering the values of the parameters $L$ and $\xi$ corresponding to $\Pi_{ext} \rightarrow 1$.

%%%%%%%%%%%%%%%%%%%%%%%%%%%%%%%%%%%%%%%%%%%%%%%%%%%%%%%%%%%%%
%%%%%%%%%%%%%%%%%%%%%%%%%%%%%%%%%%%%%%%%%%%%%%%%%%%%%%%%%%%%%

%#######################################################

\section{Conclusions}

We studied the evolution of a population embedded into a restricted space in which the interaction among the population is determined by the relative position of nodes in space. Our model generates a competition between species or individuals (represented by the nodes). Starting from a single root node and, at each time step, each existent node in the network can branch to produce up to two new daughter nodes at the next generation. The new nodes are not allowed to emerge closer than a certain distance of a pre-existent node, defined by a parameter $\xi$, i.e., overcrowding suppresses the ``fertility'' of population. Evolutionary processes are usually considered in low dimensions and, for this case, our results do not depend qualitatively on the system's dimension for $D \le 3$. 

We have demonstrated that the embedding of the network into a restricted area, which is natural for general populational evolution, set limits to growth and, for some values of the model's parameters, can result in complete extinction. The simple model we studied can schematically describe a real process in nature.

%%%%%%%%%%%%%%%%%%%%%%%%%%%%%%%%%%%%%%%%%%%%%%%%%%%%%%%%%%%%%%

\section*{Acknowledgments}

F. L. Forgerini would like to thank the FCT for the financial support by project No. SFRH/BD/68813/2010. N. C. would like to thank the Brazilian funding agencies CAPES and CNPq for the financial support. This work was partially supported by projects PTDC/FIS/108476/2008, PTDC/SAU-NEU/103904/2008, and PTDC/MAT/114515/2009.

\end{document}